# Near-infrared and Mid-infrared Light Emission of Boron-doped Crystalline Silicon

Xiaoming Wang[1#], Jiajing He[1,2*#] and Yaping Dan[1*]

[1] University of Michigan-Shanghai Jiao Tong University Joint Institute, Shanghai Jiao Tong University, Shanghai, 200240, China.

[2] Photonic Integrated Circuits Center, Shanghai Institute of Optics and Fine Mechanics, Chinese Academy of Sciences, Shanghai, 201800, China

*yaping.dan@sjtu.edu.cn; jiajinghe@siom.ac.cn

**Abstract:** The bottleneck in achieving fully integrated silicon photonics lies in silicon-based light-emitting devices that are compatible with standard CMOS technology. Dislocation loops by implanting boron into silicon and annealing represents an enticing strategy to transform highly inefficient silicon into a luminescent material. However, the emission at telecommunication wavelength suffers from the strong thermal quenching effect, resulting in low efficiency at room temperature. Here, we applied a new deep cooling process to address this issue. Interestingly, we find that electrons and holes recombine through defects emitting two photons, one in near infrared (NIR, 1.3~1.6 μm) and the other in mid-infrared band (MIR, around 3.5 μm). The PL intensity at NIR increases by three folds when the temperature increases from 77 K to 300K. Furthermore, the NIR light emission of reverse biased silicon diodes was significantly enhanced compared to forward bias, emitting the maximum output power of 42 nW at 60 mA. The results offer new opportunities for the development of IR light sources in integrated circuits.

## 1. Introduction

Electrically pumped Si based light sources are important devices for highly integrated silicon photonic systems.[1] However, achieving efficient light emission from silicon is extremely challenging due to its indirect band gap. Many approaches have been explored, including Si light emitting diodes (LED) and diodes based on rare earth-doped Si[2-4], III-V group[5], Ge-on-Si[6], and dislocation engineering[7, 8]. Dislocation loops were studied intensively as a residual damage after ion implantation doping of silicon.[9-11] However, a common issue found in these approaches is the strong thermal quenching, leading to its poor performance at room temperature (RT).[12-14] Recently, we adopted a deep cooling (DC) process to address the thermal quenching by flushing the high-temperature samples with helium gas cooled by liquid nitrogen.[4] The resultant RT photoluminescence (PL) efficiency has been remarkably improved by 2-3 orders of magnitude.

In this work, we employed the deep cooling method to treat our B-doped Si samples (2×10$^{19}$ cm$^{-3}$) and fabricated a silicon p-n junction diode by introducing additional phosphorus (P) dopants. The interstitial dislocation loops induced a localized strain, thereby facilitating the spatial confinement of injected carriers and effectively suppressing their diffusion into non-radiative pathways. Temperature-dependent photoluminescence (PL) and room-temperature electro-luminescence (EL) were recorded. Besides the broad spectrum at NIR, the emission wavelength extended to region of interest in mid-infrared (MIR) ~ 3.5 μm.

## 2. Experiments

The intrinsic <100> single crystalline silicon (FZ, resistivity > $1\times10^4$ Ω·cm) was implanted by boron ions with a dosage of $5\times10^{14}$ cm$^{-2}$ at 20 keV. The wafers were cut to pieces and then went through the same cleaning procedure in which silicon wafers were first cleaned with acetone, ethanol, deionized water, and then immersed in Piranha solution (98% Sulfuric acid: 30% Hydrogen peroxide = 3:1) to remove the contaminations. Rapid Thermal Annealing (RTA) and Deep Cooling (DC) processes were employed to treat two pieces separately to activate boron ions and repair the damage caused by ion implantation. The two samples were hold at 950 °C for 5 mins. Then, the RTA-treated sample is cooled by nitrogen gas flow and water at the same time, which takes a few minutes to drop from 950 °C to room temperature. In contrast, samples treated by the DC process was flushed with high purity Helium (99.999 %) gas cooled by liquid nitrogen (77K), causing the high-temperature (950 °C) sample to drop to -125 °C in 5 s and then rise to room temperature slowly.

The vertical PN junction diode was formed by implanting boron and phosphorus ions with a dosage of $1.5 \times 10^{14}$ cm$^{-2}$ at 20 keV and $1 \times 10^{15}$ cm$^{-2}$ at 200 keV, respectively. Deep cooling process was applied to activate the dopants. The defined region of wafers was etched by 260 nm to expose the n-type contact region. A vertical PN junction diode was formed with a pair of co-axis metal electrodes of 5 nm Cr and 100 nm Au by thermal evaporation. The cross area of emission is $2.8\times10^{-4}$ cm$^{-2}$.

The PL at NIR was excited by a 405 nm laser spot with a diameter of 2 mm. With a relatively low excitation power of 50 mW, the PL spectra were measured by a Fourier transform infrared (FTIR) spectrometer equipped with a liquid-nitrogen-cooled Ge detector system at room temperature. The EL were recorded using an Edinburgh FLS1000 Spectrometer with a nitrogen-cooled near-infrared InGaAs photomultiplier tube. The current density–voltage (J–V) characteristics of the device was measured by a Keithley 2400 sourcemeter. The emission power of the DC-treated LED was calculated with a commercial LED (Hamamatsu, L12509-0155G, 1550 nm) with a calibrated External Quantum Efficiency (EQE) as a reference.

The PL at MIR was excited by a 532 nm laser with a power of 100 mW. The PL spectra were recorded using a FTIR spectrometer (VERTEX 80v). Higher SNR can be obviously achieved with a lock-in amplifier at 1317 Hz chopper frequency. The sample was mounted on the cold head of Helium closed cycle cryostat, which allows for temperature adjustment ranging from 4 K to 300 K.

## 3. Results and Discussion

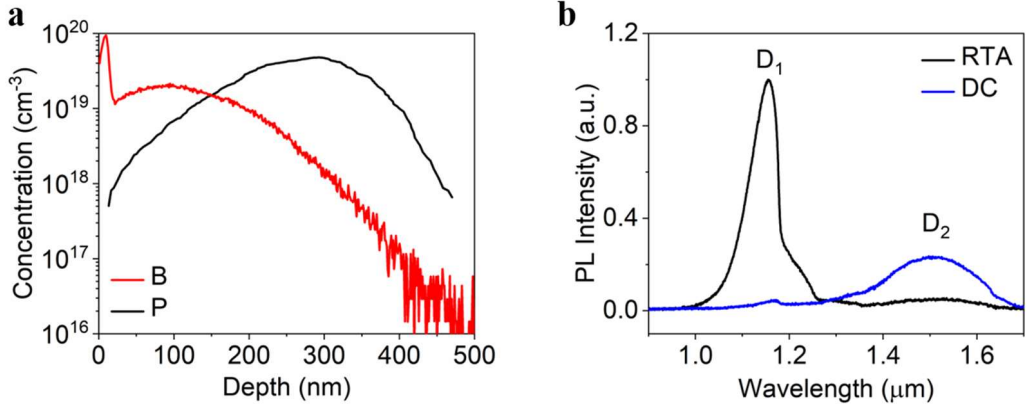

**Figure 1. (1)** Depth profile of B concentration in RTA-treated sample by SIMS. SRIM Simulation profile of P (black) concentration. **(2)** Comparison of room temperature PL spectra of the RTA-treated (black) and DC-treated (blue, purple) B-doped Si samples.

The B atom distribution profile of the sample treated by RTA has a peak concentration of $2\times10^{19}$ cm$^{-3}$ at 100 nm with a diffusion depth of 400nm according to the secondary ion mass spectroscopy (SIMS) data shown in **Fig.1a**. The kink near the surface is likely because of the accumulation of boron atoms at SiO$_2$/Si interface. As shown in **Fig.1b**, the RTA-treated B-doped Si sample has a strong peak at 1147 nm (D$_1$ line) with the excitation of 50 mW, which is related to the impurity state energy level of boron. The high doping concentration of boron makes the boron energy level continuous and become part of the valence band. However, for the same doping sample after treated with DC process, the emission of D$_1$ line is strongly suppressed and broad-band spectrum at communication band (D$_2$ line, 1.3~1.6 μm) is enhanced. This phenomenon indicates some radiative deep level defects were created by the DC process. According to semiconductor physics, electrons and holes will recombine more efficiently via deep level defects. This explains why the D$_1$ peak is suppressed and a large peak near the communication wavelength shows up. It was also previously reported that the high concentration of boron induces dislocation loops in silicon, suppressing band-to-band radiative recombination.[15,16,17] As the excitation power increases to 150 mW, both D$_1$ and D$_2$ are detectable although there is competition between the band-to-band recombination and the recombination through defects (shown in **Fig 2.a**). The Hall effect measurements show that the concentration of holes is $1\times10^{19}$ cm$^{-3}$ with the activation rate of ~50%.

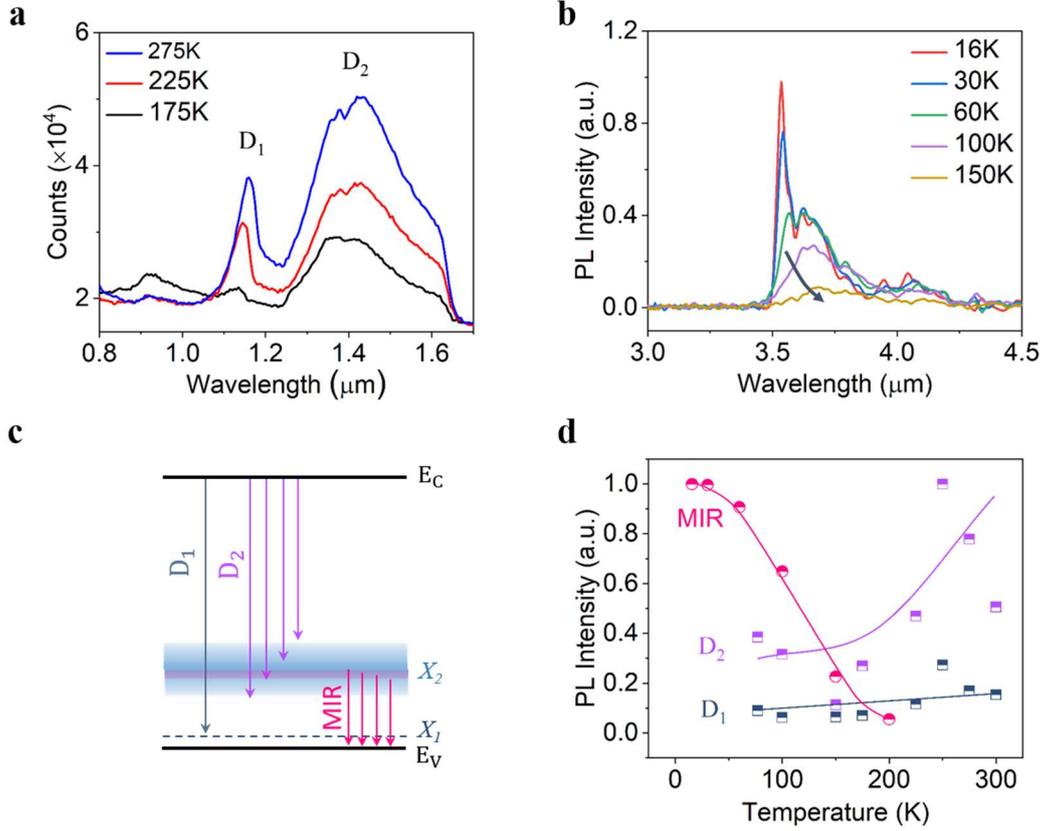

**Figure 2. Temperature-dependent PL characteristic at NIR and MIR of the DC-treated B-doped Si sample. (a)** PL spectra of NIR at 175K, 225K and 275K. **(b)** PL spectra of MIR with temperature. The solid lines shows the change of peak position as temperature. **(c)** Diagram of defect energy levels of boron in silicon. **(d)** Integrated intensity of $D_1$, $D_2$ and MIR lines with temperature. The solid lines are to guide the eye.

Temperature-dependent PL spectra were measured to investigate the luminous mechanism. **Fig.2a** exhibits the NIR spectra. The peak $D_1$ originates from the recombination of excited electrons in conduction band via the unoccupied B acceptors, generating photons with an energy near $E_g$-0.045eV ($X_1$ in **Fig.2c**). The DC process introduced quasi-continuous energy band into silicon bandgap, which formed radiative recombination centers and contributed to the $D_2$ emission bump (~ 0.86eV).

Interestingly, we found that the DC treated B-doped Si emits light at MIR shown in **Fig.2b**, in addition to the NIR emission. The MIR emission has a peak of 3.5 μm at 16 K with a red shift with increase of temperature was observed. As the temperature goes up to 150 K, the PL intensity is quenched to a level comparable to noises. The photon energy of the MIR emission is 0.31 ~ 0.35 eV. The summation of $D_2$ and MIR emission in photo energy is approximately equal to the bandgap of silicon, indicating that the MIR emission comes from the relaxation of holes from the Si valence band to the $X_2$ band.

Clearly, the DC processed B-Si samples have recombination paths depicted in **Fig.2c**. The laser excited electrons relax to the valence band through the conduction band bottom to the energy level of boron dopants ($X_1$), emitting photons near 1.08 eV ($D_1$). Another competing path is to relax excited electrons and holes to deep level defects ($X_2$ band) at the same time. Since these deep level defects distribute in a relatively wide band in the Si bandgap, the emitting photons forms a broad band in NIR ($D_2$). Holes excited from the valence band can recombine with electrons trapped in the $X_2$ band, emitting photons in MIR.

The temperature-dependent integrated intensities of $D_1$, $D_2$ (150 mW) and MIR lines were explored, shown in **Fig.2d.** The maximum values of NIR and MIR were normalized. The high concentration of boron in silicon makes the impurity-related energy levels $X_1$ degenerate. Thus, the integrated intensity of $D_1$ was almost independent with temperature. The $X_2$ energy band is deep level defect recombination centers. Decrease of temperature will reduce the recombination rate via defect recombination centers. This explains why the $D_2$ band intensity decreases at lower temperature. The MIR emission rapidly increases at lower temperature probably because the deep level defects have a stronger coupling to phonons in Si. A lower temperature will reduce the density of phonons and increase the radiative transition between the $X_2$ band and valence band.

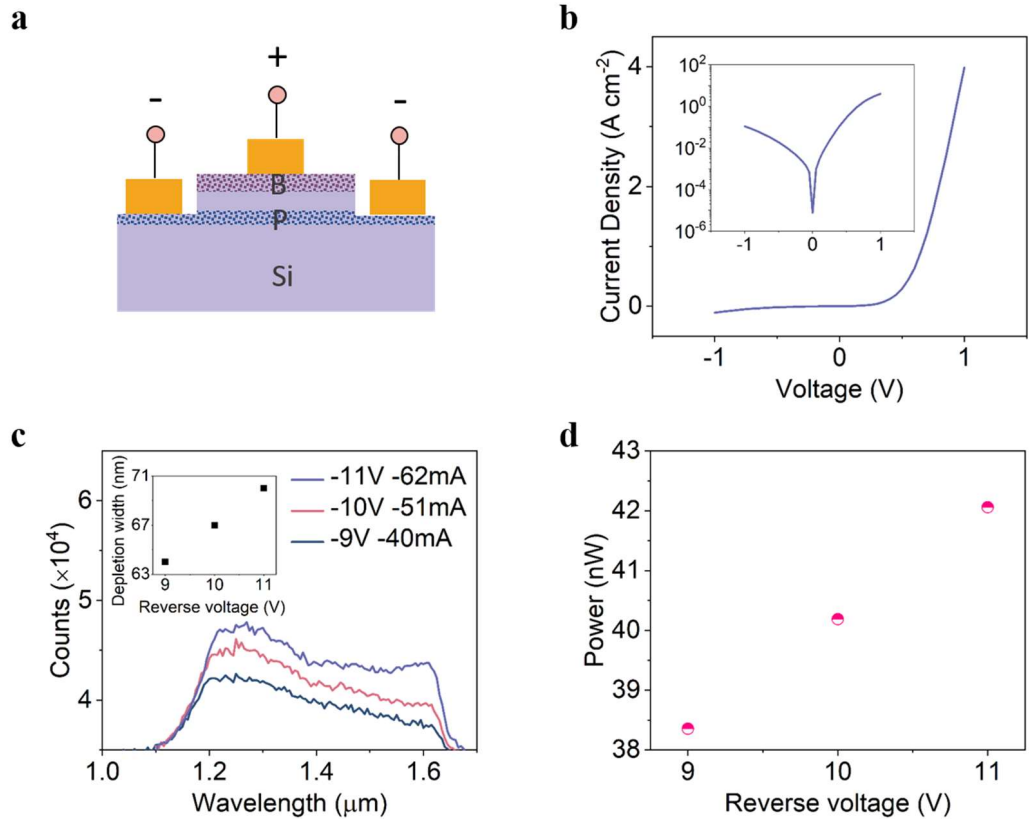

**Figure 3. B: Si LED device. (a)** Schematic of the device. **(b)** J-V curve. Inset: logarithm current vs bias. **(c)** EL spectrum under different injection currents at room temperature. **(d)** Emission power under different injection current at room temperature.

With the above characterizations of photo emission from the B-doped Si, we turned the Boron-doped Si into a vertical PN junction by pre-doping Si with phosphorus at deep location, as shown in **Fig.3a**. The SRIM simulation result indicates the distribution P atom has a peak concentration of $2\times10^{19}$ cm$^{-3}$ at 300 nm (**Fig.1a**). Metal contacts were deposited by photolithography and metal evaporation after part of the Si was removed to expose the p+ Si. The current density vs bias voltage (J-V) curve exhibits a rectifying behavior of a typical PN junction diode with an on/off ratio of ~ 60

(**Fig.3b**).

The diode has a rather weak luminescence under forward bias, comparable to the noise level. Interestingly, the luminescence becomes orders of magnitude stronger under reverse bias. The EL exhibits a broad band at room temperature ranging from 1.1 μm to a wavelength longer than 1.65 μm (limited by the InGaAs detector) as shown in **Fig.3c**. The lower end of the spectrum is limited by the silicon bandgap. Silvaco device simulations indicate that the depletion region increases from 63 nm in width at 9 V to 71 nm at 11 V. Similarly, the photo emission power from the device also linearly increases with the voltage. Note the maximum electric intensity remains ~ $3.1 – 3.4 \times 10^6$ V/cm, meaning that the electrons in the depletion region have already reached velocity saturation. The observation that the EL emission is much stronger at reverse bias than forward bias is likely caused by the impact excitation of hot electrons that are accelerated by strong electric field in the depletion region of the diode. The output power was calibrated to be ~ 38.4 – 42.3 nW (**Fig.3d**).

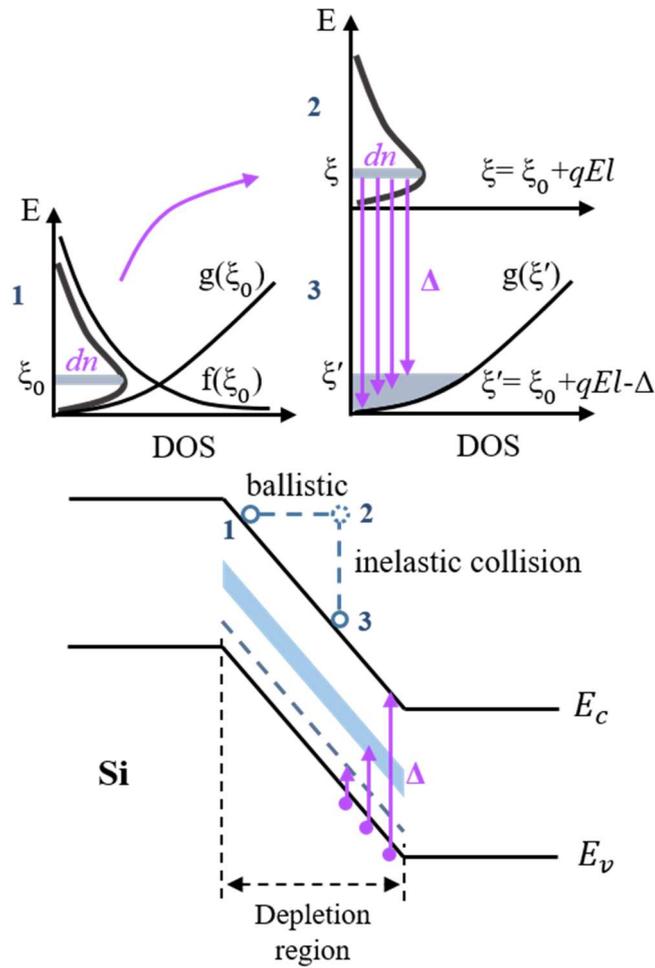

**Figure 4.** Diagram of impact excitation.

The process of impaction excitation is depicted in **Fig.4**. A PN junction diode under reverse bias will create a strong energy band bending in the depletion region. The electron and holes in the leakage current will be accelerated by the electric field. Let us use electrons as an example. Starting from the location 1, electrons will gain some kinetic energy from the electric field by ballistic transport within a distance of the electron mean free path ($l_e \approx 4\ nm$ in our case). The gained kinetic energy $qEl_e$ is estimated as 1.24 – 1.32 eV before the electrons inelastically collide impurity ions at the location 2. Since the time for electrons to ballistically transport is on an order of tens of fs, the distribution of electrons in k space has not enough time to redistribute. Therefore, it is reasonable to assume the distribution of electrons in the location 1 and 2 are the same. After collision, electrons will relax to lower energy levels at the location 3, generating a variety of quantized energies. These quantized energies will mostly likely be absorbed by electrons in the valence band, jumping to defect states in the bandgap or even in the conduction band. These excited electrons will relax and radiatively emit photons, creating a broad band EL spectrum as shown in **Fig. 3c**. The EL spectrum is different from the PL likely because the energy distribution of excitation source is different. For EL, the relaxation of hot electrons by inelastic collision will create a wide spectrum of quantized energies due to the Boltzmann distribution of electrons and because electrons in the location 2 can relax to different empty energy levels in the location 3. Therefore, electrons in the valence band will be excited to defects with different energy levels in the bandgap. For PL, optically excited electrons from the valence to conduction band will first relax to the conduction bottom and then recombine through different paths to the valence band. According to semiconductor physics, the energy level of these paths will determine the recombination efficiency of electrons at the conduction band bottom. The PL emission spectrum is thus different from the one by impact excitation.

## 4. Conclusion

We fabricated a B-doped Si light emitting diode that emit photons with a broad band from 1.1μm up to mid-infrared spectrum (~3.5 μm). Interestingly, we observed the near infrared and mid-infrared emission are coming from electrons and holes recombining through deep level defects, emitting two photons with one in near infrared and the other in mid-infrared band. The emissions from the diode are only detectable when the diode is under reverse bias. A model of impact excitation is proposed to explain this relatively strong emission under reverse bias and the spectral difference between the EL and PL.

**Acknowledgement**

X.M.W, and J.J.H contributed equally to this work. This work was supported by the National Science Foundation of China (62305354, 92065103), Shanghai Sailing Program (No. 23YF1454100), the special-key project of Innovation Program of Shanghai Municipal Education Commission (No. 2019-07-00-02-E00075), Oceanic Inter-disciplinary Program of Shanghai Jiao Tong University (SL2022ZD107), Shanghai Jiao Tong University Scientific and Technological Innovation Funds (2020QY05) and Shanghai Pujiang Program (22PJ1408200). The authors are grateful to the support for PL analysis by Man Wang at Shanghai Institute of Technical Physics, Chinese Academy of Sciences and Dr. Ruibin Wang at Instrumental Analysis Center of Shanghai Jiao Tong University. The devices were fabricated at the Center of Advanced Electronic Materials and Devices (AEMD) Shanghai Jiao Tong University.

**Data Availability Statement**

The data that support the findings of this study are available from the corresponding author upon reasonable request.